\documentclass[preprint,12pt]{aastex}

\shorttitle{Forming an O Star vis Disk Accretion?}
\shortauthors{Qiu et al.}
\begin{document}

\title{Forming an O Star via Disk Accretion?}

\author{Keping Qiu\altaffilmark{1,2}, Qizhou Zhang\altaffilmark{3}, Henrik Beuther\altaffilmark{4}, Cassandra Fallscheer\altaffilmark{4,5}}\email{kqiu@mpifr-bonn.mpg.de}
\altaffiltext{1}{Max-Planck-Institut f\"{u}r Radioastronomie, Auf dem H\"{u}gel 69, 53121 Bonn, Germany}
\altaffiltext{2}{Key Laboratory of Modern Astronomy and Astrophysics (Nanjing University), Ministry of Education, Nanjing 210093, China}
\altaffiltext{3}{Harvard-Smithsonian Center for Astrophysics, 60 Garden Street, Cambridge, Massachusetts 02138, USA}
\altaffiltext{4}{Max-Planck-Institut f\"{u}r Astronomie, K\"{o}nigstuhl 17, 69117 Heidelberg, Germany}
\altaffiltext{5}{Department of Physics and Astronomy, University of Victoria, 3800 Finnerty Road, Victoria, BC V8P 5C2, Canada}

\begin{abstract}
We present a study of outflow, infall, and rotation in a $\sim$10$^5~L_{\odot}$
star-forming region, IRAS 18360-0537, with Submillimeter Array (SMA) and IRAM 30m
observations. The 1.3 mm continuum map shows a 0.5 pc dust ridge, of which the
central compact part has a mass of $\sim$80 $M_{\odot}$ and harbors two
condensations, MM1 and MM2. The CO (2--1) and SiO (5--4) maps reveal a biconical
outflow centered at MM1, which is a hot molecular core (HMC) with a gas temperature
of $320\pm50$ K and a mass of $\sim$13 $M_{\odot}$. The outflow has a gas mass of
$54~M_{\odot}$ and a dynamical timescale of $8\times10^3$ yr. The kinematics of the
HMC is probed by high-excitation CH$_3$OH and CH$_3$CN lines, which are detected at
sub-arcsecond resolution and unveil a velocity gradient perpendicular to the outflow
axis, suggesting a disk-like rotation of the HMC. An infalling envelope around the
HMC is evidenced by CN lines exhibiting a profound inverse P-Cygni profile, and the
estimated mass infall rate, $1.5\times10^{-3}~M_{\odot}$\,yr$^{-1}$, is well
comparable to that inferred from the mass outflow rate. A more detailed investigation
of the kinematics of the dense gas around the HMC is obtained from the $^{13}$CO and
C$^{18}$O (2--1) lines; the position-velocity diagrams of the two lines are
consistent with the model of a free-falling and Keplerian-like rotating envelope. The
observations suggest that the protostar of a current mass $\sim$10 $M_{\odot}$
embedded within MM1 will develop into an O star via disk accretion and envelope
infall.
\end{abstract}

\keywords{stars: formation --- stars: early-type --- ISM: kinematics and dynamics --- ISM: jets and outflows}

\section{Introduction} \label{intro}
The formation process of high-mass stars ($>8~M_{\odot}$) is poorly understood
\citep{Zinnecker07}. One key question in the field is whether high-mass stars form as
a scaled-up version of low-mass star formation \citep[e.g.,][]{Keto10,Johnston11}. In
a standard paradigm of low-mass star formation, a protostar accretes from a
circumstellar disk embedded within an infalling envelope \citep{Shu87}. Since 
accretion disks and collimated outflows are physically connected, the detection of 
well collimated outflows in an increasing number of high-mass star-forming regions
supports an accretion-based scenario for massive star formation
\citep[e.g.,][]{Beuther02,Zhang07a,Zhang07b,Qiu07,Qiu09a,Qiu11a,Zapata11}.
In the meantime, great efforts have been made in searching for direct evidence for an
infalling envelope and a rotating disk in association with massive star formation.
Nevertheless, only a very few massive star-forming cores have been found to show
clear signatures of both infall and rotation. Beltr\'{a}n (2011) enlists a sample of
six such sources from the literature, G10.62-0.38, G19.61-0.23, G24.78+0.08 A1,
G31.41+0.31, W51 e2, and W51 North. But some more cases exist \citep[e.g., NGC 7538
IRS 1,][]{Qiu11b,Beuther12}. In these studies, a rotation is inferred from a velocity
gradient seen in high-density tracing molecular lines and an infall is probed by
molecular lines with redshifted absorption against a bright continuum source. Even
this small sample should be taken with caution. For example, the orientations of the
velocity gradients may change by tens of degrees, depending on the lines being
analyzed \citep[e.g., W51 e2:][]{Zhang97,Zhang98,Keto08,Klaassen09}. Meanwhile, if a
velocity gradient is arising from a rotating disk, it is expected to be perpendicular
to a bipolar outflow, but for most sources in the sample, such an association between
a putative disk and an outflow is yet to be confirmed
\citep[e.g., G31.41+0.31,][and references therein]{Cesaroni11}. Therefore,
identifying and characterizing clear-cut examples showing self-consistent outflow,
infall, and rotation motions remains of great interest to our understanding of
massive star formation. Here we present such a study toward IRAS 18360-0537, using
Submillimeter Array\footnote[6]{The SMA is joint project between the Smithsonian
Astrophysical Observatory and the Academia Sinica Institute of Astronomy and
Astrophysics and is funded by the Smithsonian Institution and the Academia Sinica.}
(SMA) $0.\!''5$--$1.\!''5$ resolution observations in the 1.3 mm waveband.

IRAS 18360-0537 has a kinematic distance of 6.3 kpc and a far-IR luminosity of
$1.2\times10^5~L_{\odot}$ \citep{Molinari96}. In earlier single-dish surveys, the
region was found to be associated with water maser emission \citep{Palla91,Brand94}
and thermal NH$_3$ \citep{Molinari96} and CS \citep{Bronfman96} emission. There is so
far no report on high-angular-resolution observations of the region in
(sub)millimeter or centimeter wavelengths.

\section{Observations} \label{obs}
The SMA observations were undertaken in two frequency setups, each with two array
configurations. The observing campaign was designed to cover CO, $^{13}$CO,
C$^{18}$O (2--1), and several CH$_3$OH and CH$_3$CN lines in one frequency setup and
to cover CN ($N$=2--1) and SiO (5--4) lines in the other setup. The CO and SiO lines
are typical outflow tracers, while the CN lines may reveal potential infall
signatures in high-mass star-forming cores \citep{Zapata08,Wu09}. The kinematics of
the innermost dense gas can be inferred from high-excitation CH$_3$OH and CH$_3$CN
lines. The details of the observations are presented in Table \ref{table1}.

\begin{figure}
\epsscale{.9} \plotone{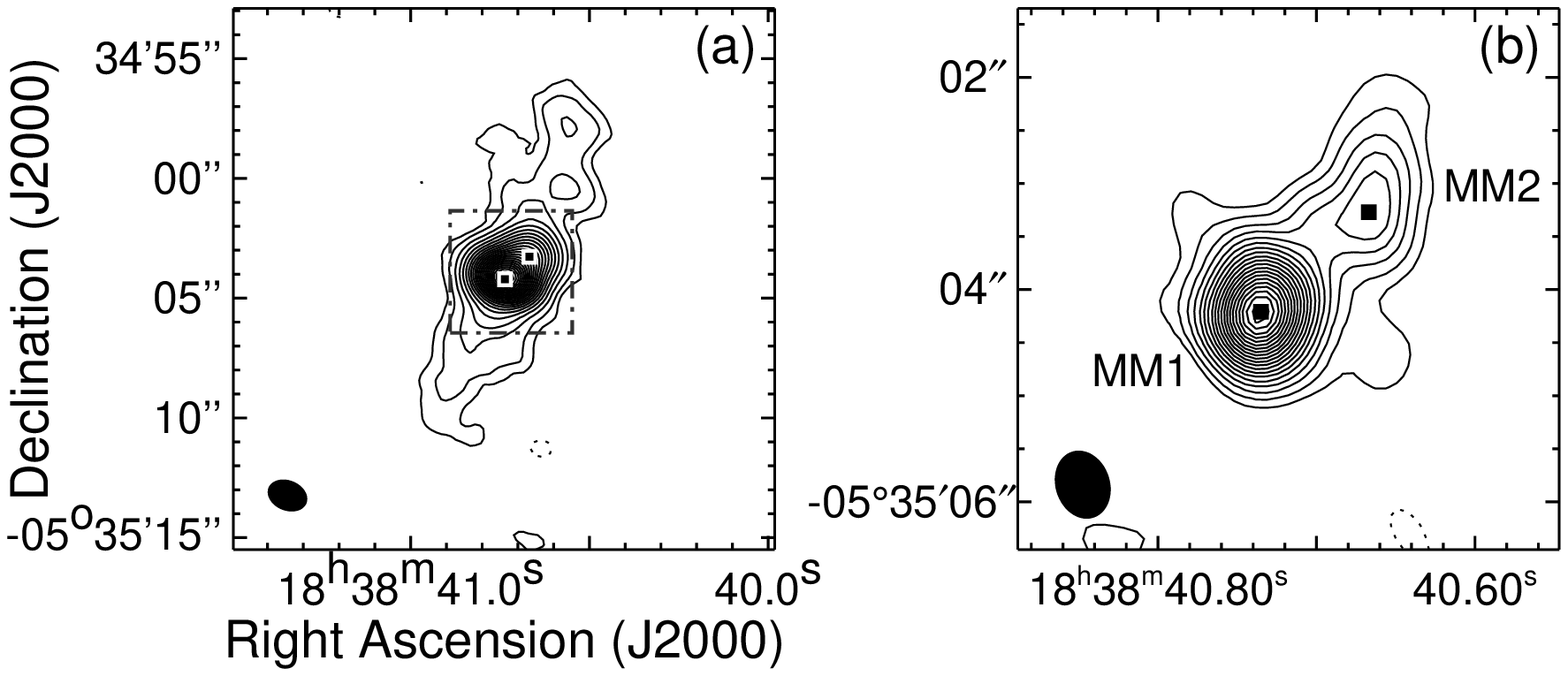} \caption{(a) 1.3 mm continuum map constructed by combining data from all the observations. Solid and dotted contours show positive and negative emission, respectively, with the contour levels of  $6\times(1,2,3,...,23)^{1.5}$ mJy\,beam$^{-1}$. (b) Same as (a), but for the EXT and VEX data, and contour levels of $9\times(1,2,3,...,21)^{1.3}$ mJy\,beam$^{-1}$. In each panel (and hereafter), two filled squares denote the peak positions of the two condensations (MM1 and MM2), and a filled ellipse in the lower left shows the synthesized beam.
\label{cont}}
\end{figure}

The data were calibrated using the IDL MIR package, and imaged with the MIRIAD
software. A continuum data set was constructed from line free channels using the
MIRIAD task UVLIN. We performed self-calibration on the continuum, and applied
the solutions to both continuum and line data sets. For molecular line maps within
each frequency setup, the calibrated visibilities from the two array configurations
were jointly imaged. The angular resolution (the synthesized beam,
${\theta}_{\rm syn}$, and its position angle, PA) and rms noise level, $\sigma$, of a
map vary with the visibility weighting algorithm. Most maps were made with a
compromise between resolution and sensitivity (e.g., varying the ROBUST parameter),
resulting in ${\theta}_{\rm syn}\sim2.\!''7\times1.\!''4$ at ${\rm
PA}\sim62^{\circ}$, $\sigma\sim50$ mJy\,beam$^{-1}$ per 1.5 km\,s$^{-1}$ in CO lines,
${\theta}_{\rm syn}\sim2.\!''0\times1.\!''4$ at ${\rm PA}\sim64^{\circ}$,
$\sigma\sim30$ mJy\,beam$^{-1}$ per 1.5 km\,s$^{-1}$ in SiO (5--4), and
${\theta}_{\rm syn}\sim1.\!''4\times1.\!''3$ at ${\rm PA}\sim57^{\circ}$,
$\sigma\sim35$ mJy\,beam$^{-1}$ per 0.6 km\,s$^{-1}$ in CN lines. To achieve the
highest possible angular resolution, which is desirable in searching for disk-like
structures, the CH$_3$OH and CH$_3$CN lines were made with a uniform weighting,
leading to ${\theta}_{\rm syn}\sim0.\!''8\times0.\!''6$ at ${\rm PA}\sim20^{\circ}$,
$\sigma\sim65$ mJy\,beam$^{-1}$ per 0.6 km\,s$^{-1}$. For the 1.3 mm continuum
emission, we made a map combining visibilities from all four array configurations
(Figure \ref{cont}(a)), with ${\theta}_{\rm syn}\sim1.\!''6\times1.\!''2$ at ${\rm
PA}\sim68^{\circ}$, $\sigma\sim1.5$ mJy\,beam$^{-1}$, and a map combining the EXT and
VEX data (Figure \ref{cont}(b)), with ${\theta}_{\rm syn}\sim0.\!''6\times0.\!''5$ at
${\rm PA}\sim18^{\circ}$, and $\sigma\sim2.7$ mJy\,beam$^{-1}$.

The IRAM\footnote[7]{IRAM is supported by INSU/CNRS (France), MPG (Germany) and IGN
(Spain).} 30m telescope was used to fill the zero-spacing of the SMA maps in CO and
$^{13}$CO (2--1). The observations were conducted on 2007 November 1. A $2'\times2'$
region centered at (R.A.,
Dec.)$_{\mathrm J2000}$=($18^{\mathrm h}38^{\mathrm m}40.\!^{\mathrm s}3,
-5^{\circ}35{'}6.\!{''}0$) was mapped in the on-the-fly mode. Scans were made in both
Right Ascension and Declination directions in an effort to prevent systematic
scanning effects from appearing in the data. The receiver was tuned to 230.538 GHz
and 220.399 GHz for the CO and $^{13}$CO (2--1) lines, respectively, and the spectra
have a 0.4 km\,s$^{-1}$ resolution. The spectra were processed using CLASS in the
GILDAS software package. The CO spectra have an rms noise level of 1.1 K and the
$^{13}$CO spectra have an rms of 0.5 K in $T_{\mathrm {mb}}$. The combination of the
SMA and single-dish data was performed in MIRIAD following a procedure outlined in
Zhang et al. (1995).

\section{Results} \label{result}
\subsection{Dust Condensations} \label{res_cont}
Figure \ref{cont}(a) shows the 1.3 mm continuum map made from observations conducted
with the four array configurations. The emission reveals a northwest-southeast
(NW-SE) ridge with a bright and compact part at the center. In Figure \ref{cont}(b),
relatively extended emission is filtered out in the map made with the EXT and VEX 
data, and the central part is resolved into two condensations, MM1 and MM2. This work
focuses on the brightest dust condensation, MM1. It shows very faint free-free 
emission ($\sim$0.5 mJy at 1--3 cm) in our Very Large Array observations (Qiu et al. 
in prep.). The observed 1.3 mm continuum emission is completely dominated by dust 
emission.

\begin{figure}
\epsscale{.9} \plotone{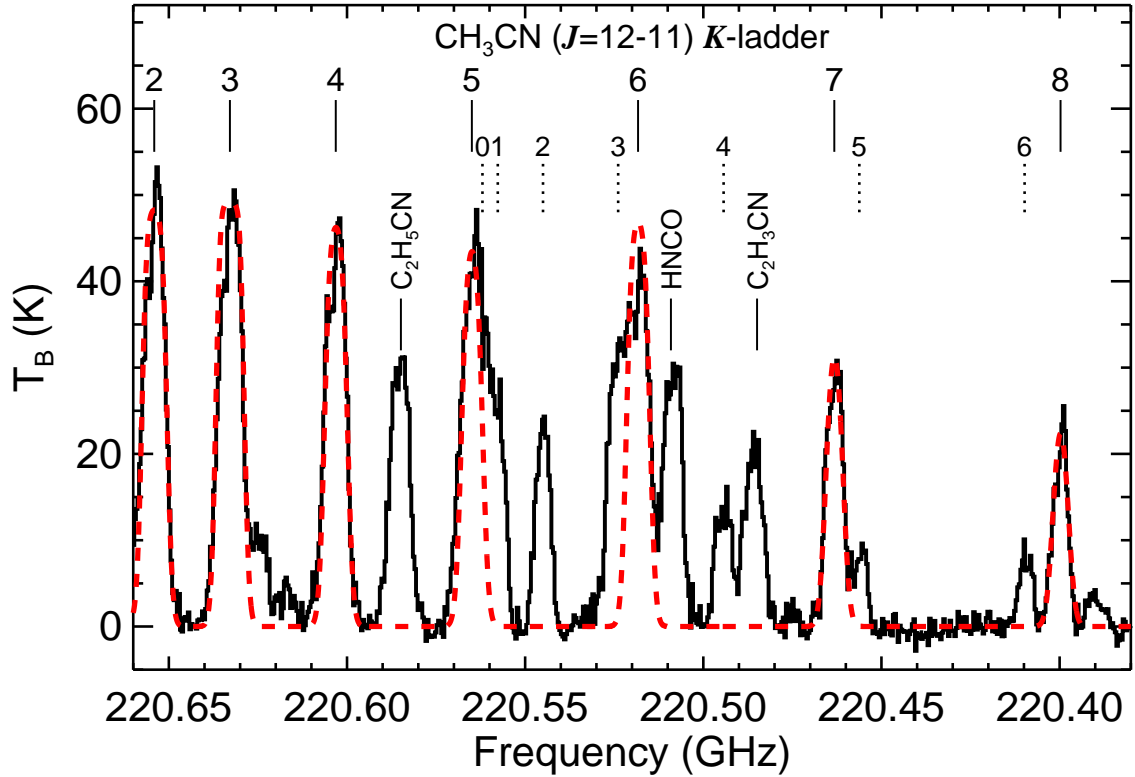} \caption{The spectra toward MM1, covering the $K$=2--8 ladder of CH$_3$CN (12--11), with the observations shown in black histograms and the best fit LTE model in dashed lines. The CH$_3$CN (12--11) $K$=2--8 components are labeled with solid bars, and the CH$_3^{13}$CN (12--11) $K$=0--6 components with dotted bars. Some other lines covered in the frequency range are also denoted.
\label{ch3cn}}
\end{figure}

MM1 shows rich emission lines in complex organic molecules, e.g., CH$_3$OH, CH$_3$CN,
C$_2$H$_3$CN, and C$_2$H$_5$CN. The CH$_3$CN (12--11) $K$-ladder is a good
thermometer of dense gas in high-mass star-forming regions. Following Qiu \& Zhang
(2009) and Qiu et al. (2011b), we simultaneously fit the $K=2$--8
components\footnote[8]{The relatively low excitation $K=0$, 1 components are resolved
out in the VEX data.}, taking into account the optical depth effect. The systemic
velocity, 103.5 km\,s$^{-1}$, is determined from Gaussian fittings to the $K=7$, 8
components, which are of the highest excitation and not heavily blended with other
lines. This determination is concurred by the C$_2$H$_5$CN
(25$_{22,4}$--24$_{22,3}$) and CH$_3^{13}$CN (12$_2$--11$_2$) lines (see Figure
\ref{ch3cn}), to which Gaussian fittings give velocities of 102.9 and 103.5
km\,s$^{-1}$, respectively. In Figure \ref{ch3cn}, the best-fit model reasonably
matches the observation, and yields a gas temperature of $320\pm50$ K, where the
uncertainty accounts for the 1$\sigma$ noise level. The fitting assumes that all the
$K$ components are tracing the same gas under local thermodynamic equilibrium
conditions. Provided potential density and temperature gradients for an internally
heated protostellar core, the best-fit model may represent a characteristic estimate.

Assuming thermal equilibrium between the dust and dense gas, the derived gas
temperature in MM1 approximates the dust temperature. To constrain the dust opacity
index, $\beta$, we make use of the 880 $\mu$m continuum data obtained from recent SMA
observations. By comparing the 1.3 mm and 880 $\mu$m continuum maps constructed with
the same $(u,v)$ range, we deduce $\beta=0.82$ toward MM1 and $\beta=0.50$ toward
MM2, corresponding to dust opacities of 2.5 and 4.3 cm$^2$\,g$^{-1}$, respectively,
at 1.3 mm \citep{Hildebrand83}. The gas mass of MM1 is then estimated from its 1.3 mm
dust continuum flux, 0.76 Jy, to be 13 $M_{\odot}$, by adopting a canonical
gas-to-dust mass ratio of 100. The mass of MM2 is less clear given its unknown gas or
dust temperature. Adopting a temperature of 30 K deduced from prior NH$_3$
observations \citep{Molinari96}, one obtains a mass of 25 $M_{\odot}$ from the
continuum flux of 0.2 Jy.

\subsection{Bipolar Outflow}
\begin{figure}
\epsscale{.45} \plotone{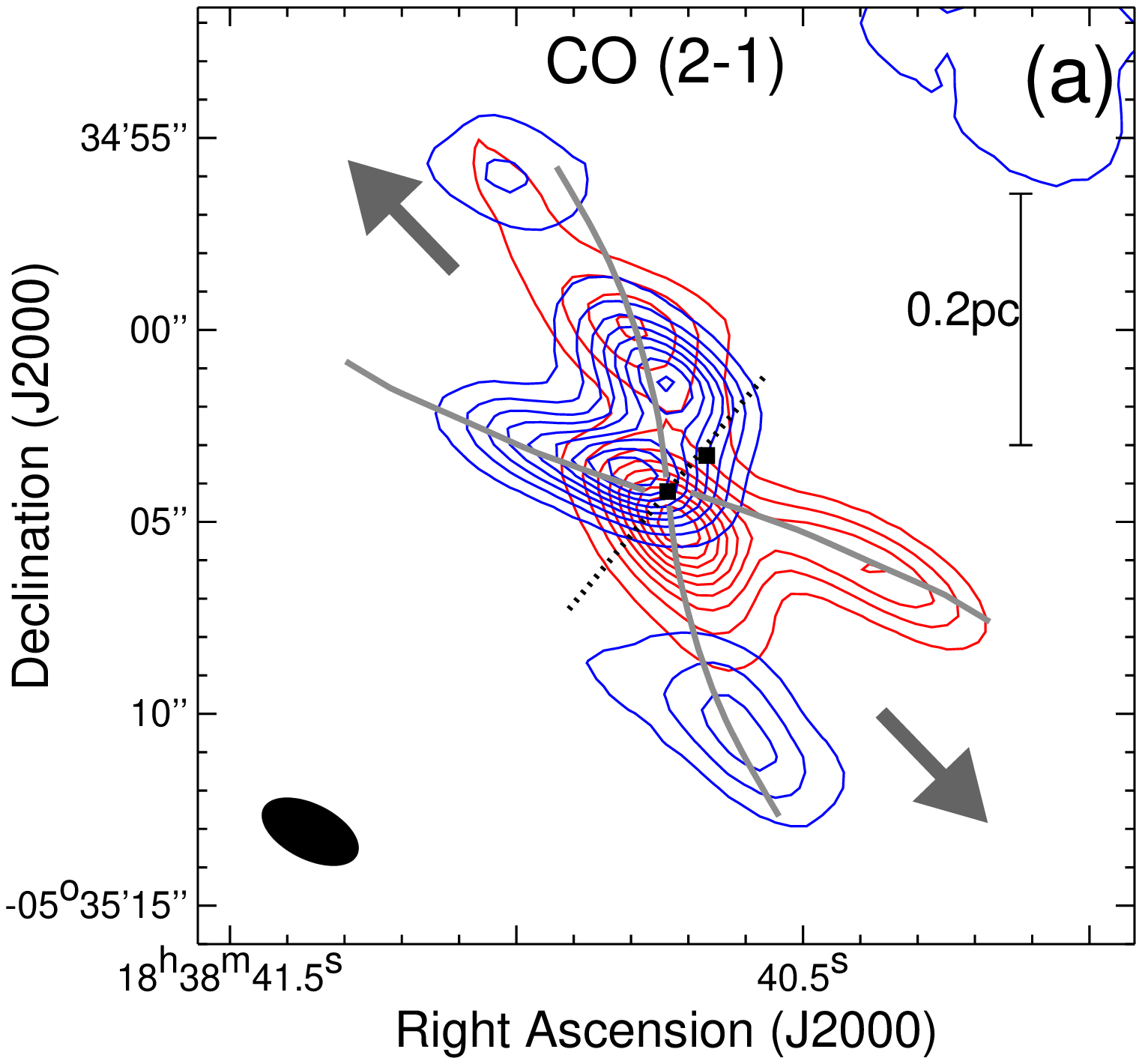}\plotone{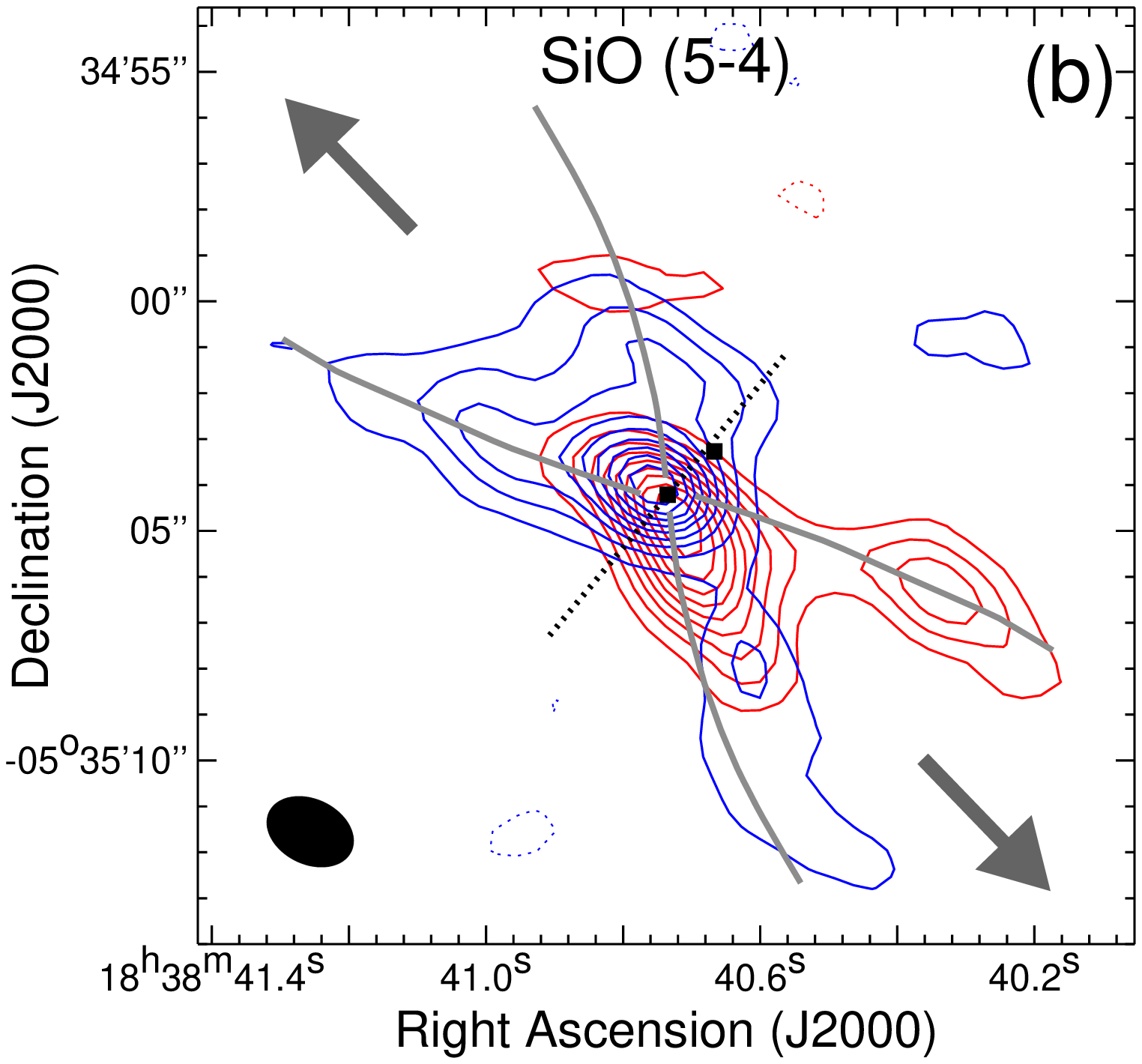} \caption{(a) CO (2--1) emission
observed with the SMA, and integrated from 69 to 99 km\,s$^{-1}$ for the blueshifted lobe and from 106.5 to 139.5 km\,s$^{-1}$ for the redshifted lobe, shown in blue and red contours, respectively. The starting and spacing contour levels are 10\% of the peak emission, which is 56.4 Jy\,beam$^{-1}$~km\,s$^{-1}$ for the blueshifted lobe and 71.3 Jy\,beam$^{-1}$~km\,s$^{-1}$ for the redshifted lobe. The gray curves outline a biconical outflow, and the two arrows indicate the orientation of the outflow axis. A dotted line marks the direction of a velocity gradient seen in MM1 (see Figure \ref{rotation}). (b) The same as (a), but for SiO (5--4). The blueshifted lobe is integrated from 88.5 to 102 km\,s$^{-1}$, and the redshifted lobe from 105 to 118.5 km\,s$^{-1}$. The starting and spacing contour levels are 10\% of the peak emission, which is 6.7 Jy\,beam$^{-1}$~km\,s$^{-1}$ for the blueshifted lobe and 7.3 Jy\,beam$^{-1}$~km\,s$^{-1}$ for the redshifted lobe.
\label{outflow}}
\end{figure}

Figure \ref{outflow} shows the velocity integrated CO (2--1) and SiO (5--4) emission
observed with the SMA. Both CO and SiO emission seems to be dominated by a
northeast-southwest (NE-SW) bipolar outflow centered at MM1. The outflow has a
biconical shape with an opening angle of $\sim$65$^{\circ}$. Such an outflow is often
thought to form from ambient gas being swept up or entrained by a wide-angle wind
\citep[e.g.,][]{Shepherd98,Qiu09b}. A blueshifted clump to the south of MM1 and a
redshifted clump to the north, which are more prominent in the CO map, could be
part of the biconical outflow, if the wide-angle wind is slightly inclined to the
plane of sky and impinges into an environment with an inhomogeneous density structure.
However, the possibility that the two clumps form another outflow cannot be ruled
out.

\begin{figure}
\epsscale{1} \plotone{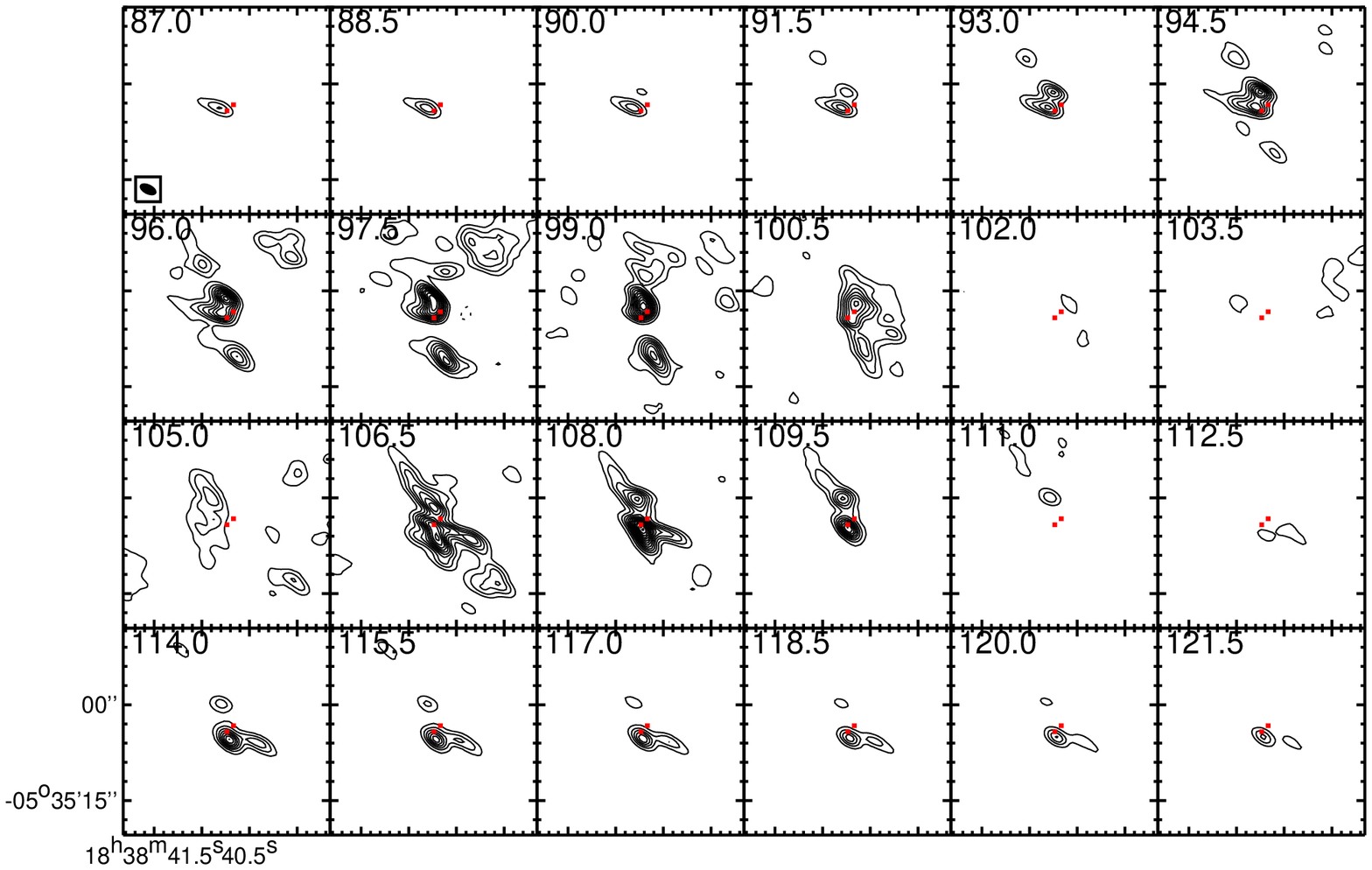}\plotone{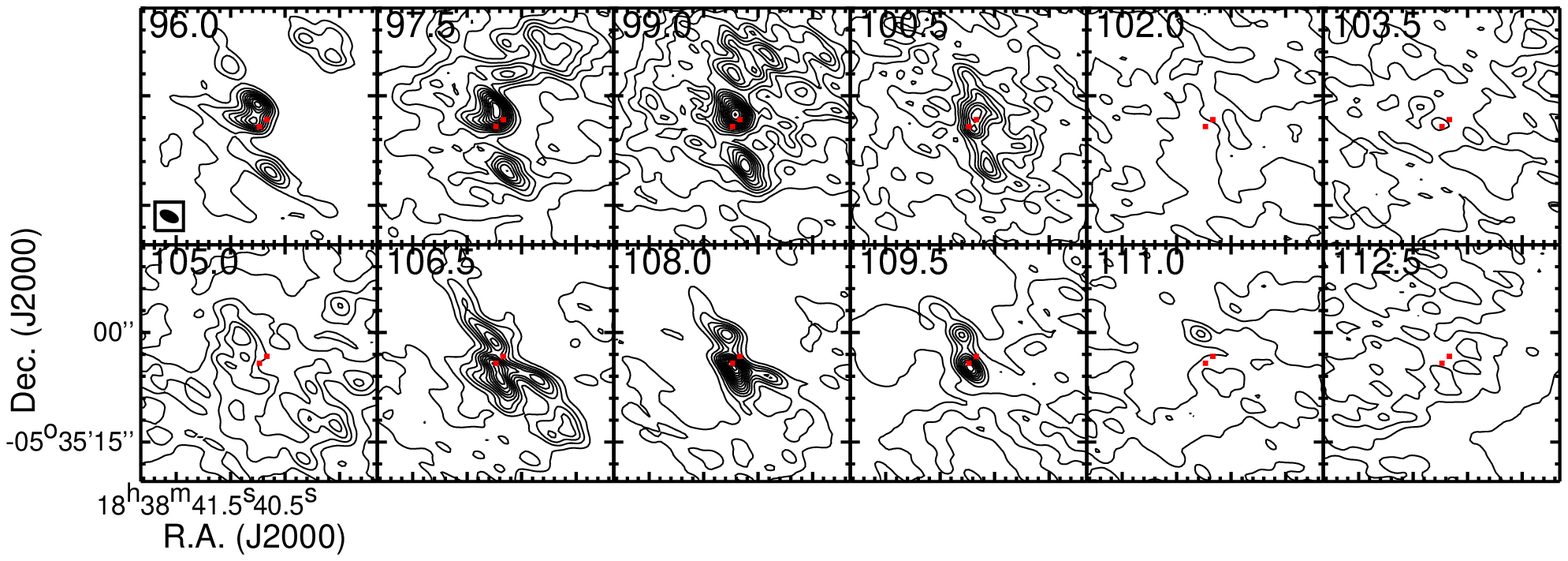} \caption{SMA CO (2--1) velocity channel maps. For channels of 96--112.5 km\,s$^{-1}$, the maps made with the combined SMA and IRAM 30m data are shown on the bottom. Solid and dotted contours show the positive and negative emission, respectively, with the same absolute levels starting from and increasing in steps of 0.6 Jy\,beam$^{-1}$.
\label{co_ch}}
\end{figure}

\begin{figure}
\epsscale{1} \plotone{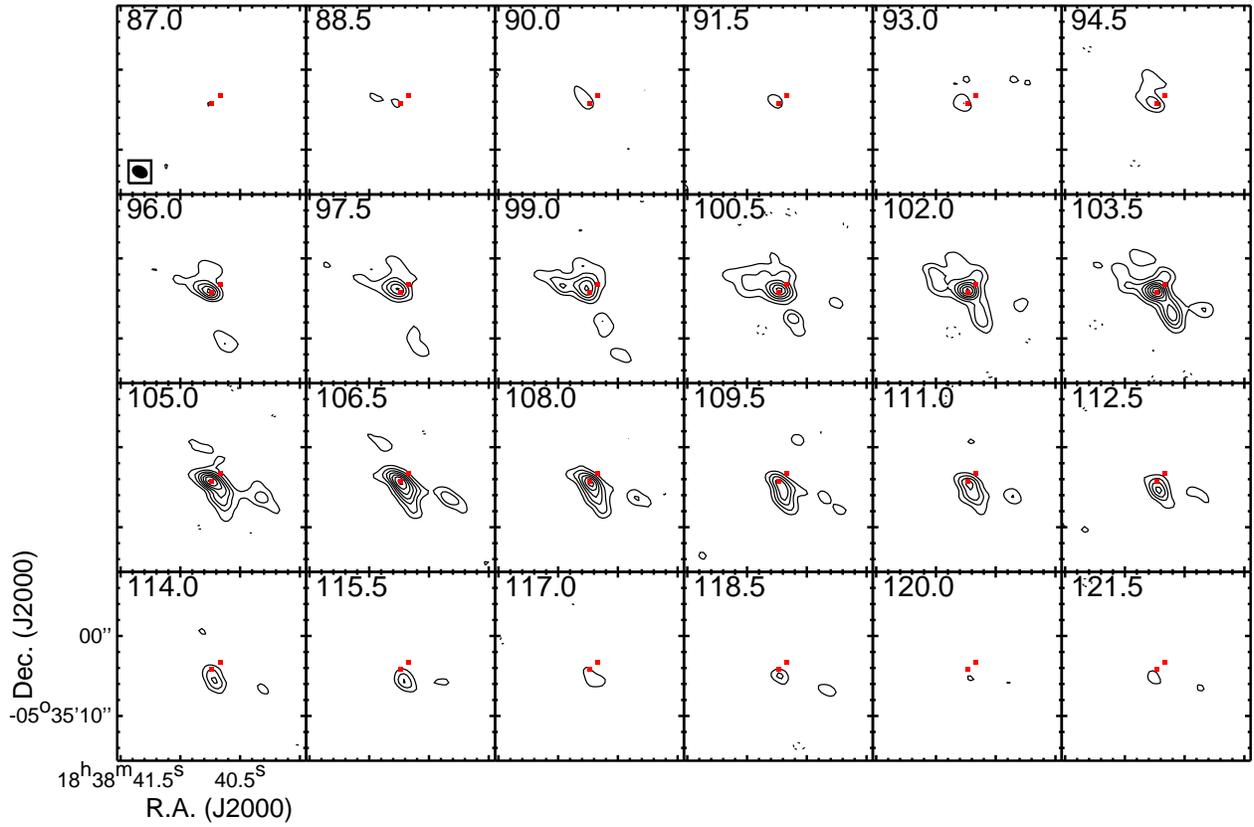} \caption{SMA SiO (5--4) velocity channel maps. Solid and dotted contours show the positive and negative emission, respectively, with the same absolute levels starting from and increasing in steps of 0.12 Jy\,beam$^{-1}$.
\label{sio_ch}}
\end{figure}

Figure \ref{co_ch} shows the velocity channel maps of the CO emission observed with
the SMA; at low velocities (96--112.5 km\,s$^{-1}$), the maps made with the combined
SMA and IRAM 30m (SMA+30m) data are also presented. It is clear that the combined
data effectively recover extended emission around the cloud velocity. Emission
arising from outflow structures appears predominant in channels of $\leq$100.5
km\,s$^{-1}$ and $\geq$106.5 km\,s$^{-1}$. In particular, the cone-shaped structure
of the SW lobe is most noticeable in the 106.5 km\,s$^{-1}$ channel map. The emission
at 111.0 and 112.5 km\,s$^{-1}$ is mostly filtered out in the SMA maps and appears to
be dominated by diffuse gas in the combined maps. This is most likely due to the
self-absorption effect caused by a foreground cloud. Figure \ref{sio_ch} shows the
SiO (5--4) channel maps. SiO is believed to be produced in the gas phase through
shock-driven sputtering of Si-bearing species on dust grains \citep{Schilke97}.
Therefore the SiO emission is not contaminated by emission from quiescent ambient
gas. As seen in the integrated map, the SiO outflow is in general consistent with the
outflow seen in the CO emission.

\begin{figure}
\epsscale{1} \plotone{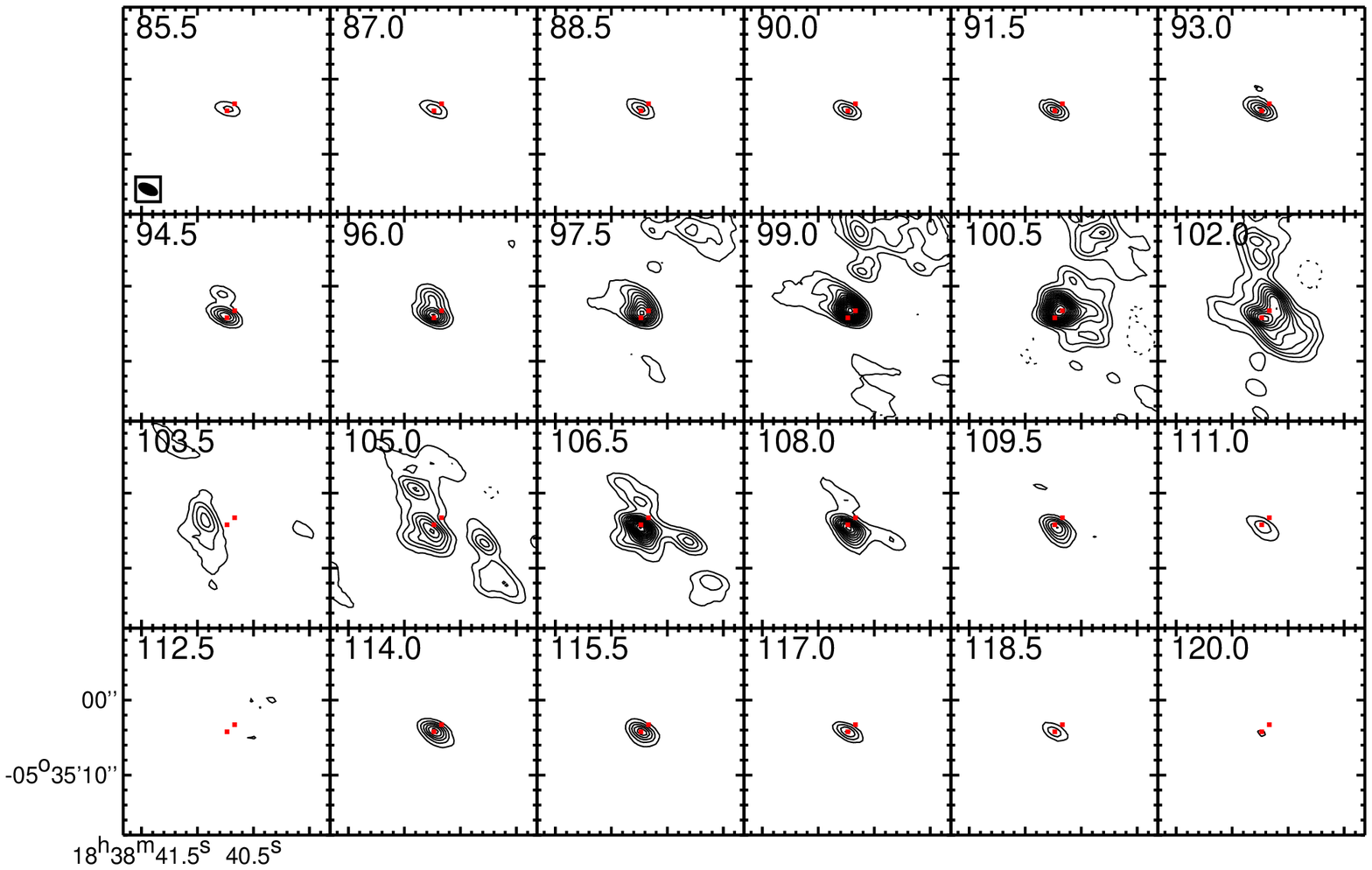}\plotone{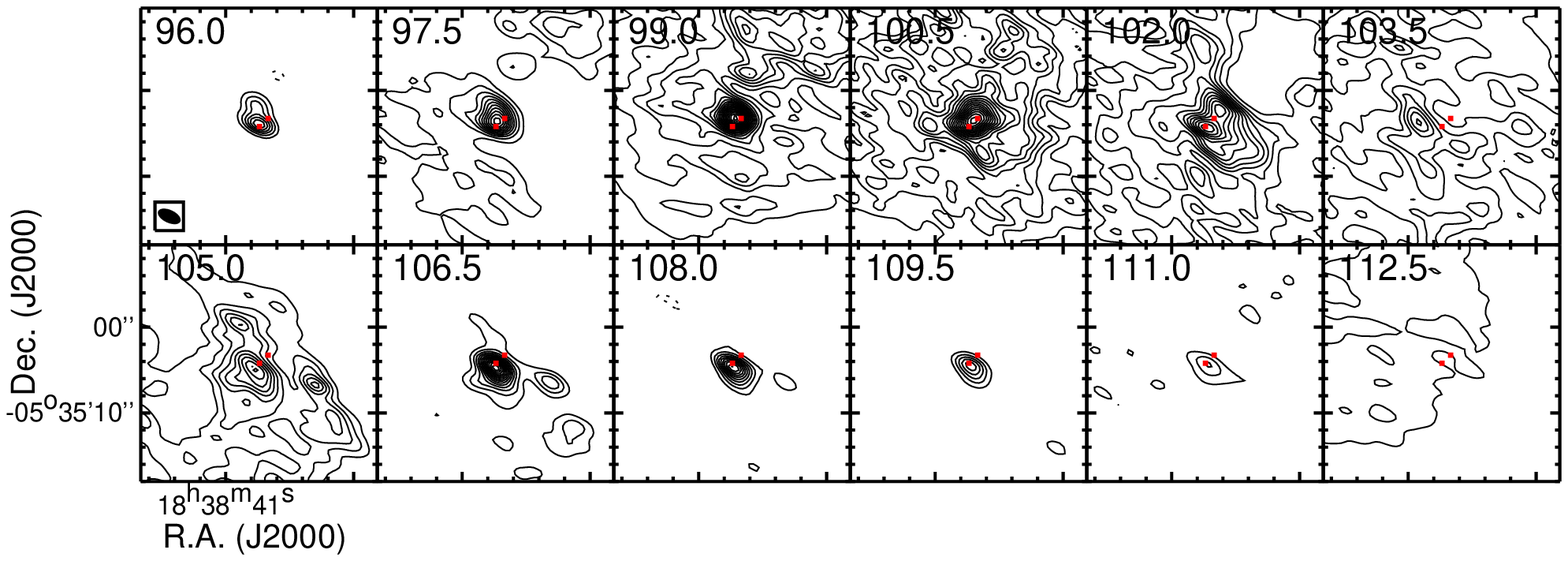} \caption{SMA $^{13}$CO (2--1) velocity channel maps. For channels of 96--112.5 km\,s$^{-1}$, the maps made with the combined SMA and IRAM 30m data are shown on the bottom. The contour levels start from and increase in steps of 0.3 Jy\,beam$^{-1}$.
\label{13co_ch}}
\end{figure}

\begin{figure}
\epsscale{.9} \plotone{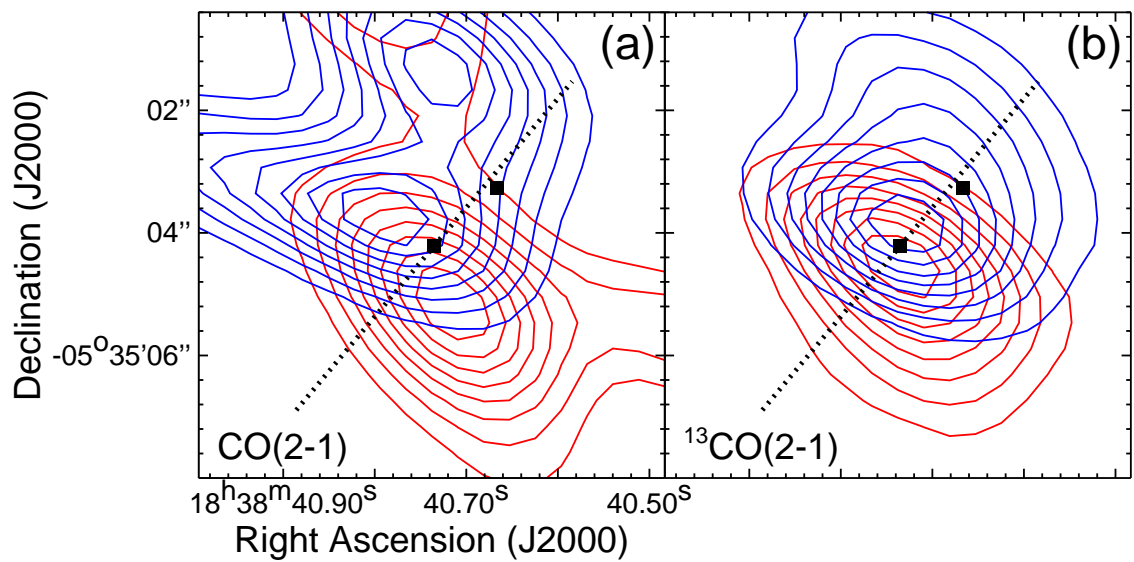} \caption{(a) A close-up of the SMA CO (2--1) velocity integrated map. Blue and red contours show emission integrated from 69 to 100.5 km\,s$^{-1}$ and from 106.5 to 139.5 km\,s$^{-1}$, respectively. The contour levels start from 20\% and increase in steps of 10\% of the peak emission, which is 59.1 and 71.3 Jy\,beam$^{-1}$~km\,s$^{-1}$ for the blue and red lobes, respectively. (b) The same as (a), but for $^{13}$CO (2--1), showing emission integrated from 85 to 102 km\,s$^{-1}$ and from 105 to 120 km\,s$^{-1}$. The peak emission is 37.7 and 26.2 Jy\,beam$^{-1}$~km\,s$^{-1}$ for the blue and red lobes, respectively.
\label{co_13co}}
\end{figure}

Figure \ref{13co_ch} shows the SMA $^{13}$CO (2--1) channel maps as well as the
SMA+30m maps at low velocities. The $^{13}$CO emission arising from the NE-SW outflow
is less extended than that in CO (2--1), but is still appreciable in the channels of
94.5--108.0 km\,s$^{-1}$ (except the 103.5~km\,s$^{-1}$ channel). In the close
vicinity of MM1, the $^{13}$CO emission peaks to the NW of MM1 at blueshifted
velocities and to the SE at redshifted velocities; the offset between the $^{13}$CO
peak and dust continuum peak decreases with the velocity so that the $^{13}$CO
emission approximately coincides with MM1 at the highest velocities. On the other
hand, the CO (2--1) emission peak is clearly offset from MM1 even at the highest
velocities, and shifts from the NE to SW. Such a difference is highlighted in Figure
\ref{co_13co}. While the relatively extended emission in $^{13}$CO traces the
outflow, the compact and bright emission closely around MM1 is predominantly arising
from a dense envelope.

\subsection{Velocity Gradients in High-excitation CH$_3$OH and CH$_3$CN Lines}
The $^{13}$CO (2--1) observations provide hints for the rotation of dense gas around
MM1 (Figure \ref{co_13co}). However, the $^{13}$CO (2--1) line is affected by the
outflow and, with a critical density, $n_{\rm cr}$, of order $10^4$ cm$^{-3}$ and an 
upper level energy, $E_{\rm up}$, around 16 K above the ground, cannot effectively 
probe the hot and dense gas within MM1. To search for a potential disk-like 
structure, which is supposed to lie in the innermost part of a molecular core, we 
inspected high-excitation lines from less abundant molecular species. A NW-SE 
velocity gradient across MM1 is consistently seen. Figure \ref{rotation} shows the 
zeroth moment (velocity integrated emission) and first moment (intensity weighted 
velocity) maps, and position-velocity ($P$-$V$) diagrams of the CH$_3$OH
(15$_{4,11}$--16$_{3,13}$)$E$, CH$_3$CN ($12_7$--$11_7$), and vibrational CH$_3$CN
($12_1$--$11_1$) lines. These lines are chosen for the following reasons. First,
they are highly excited ($E_{\rm up}>350$ K) and, as evidenced by the zeroth moment
maps, probe the innermost dense gas in MM1. Second, they are not significantly
blended with other lines; CH$_3$CN ($12_7$--$11_7$) is slightly blended with
CH$_3^{13}$CN ($12_5$--$11_5$), but the two lines are $\sim$9.5 km\,s$^{-1}$ apart
and the former is three times brighter than the latter, so it is feasible to ignore
the contribution from CH$_3^{13}$CN ($12_5$--$11_5$) in interpreting CH$_3$CN
($12_7$--$11_7$). Finally, these lines are detected with sufficient signal-to-noise
ratios in the VEX observations so that the velocity gradient is discernable at a
$0.\!''4$ resolution; some other lines with a higher $E_{\rm up}$, e.g., CH$_3$CN
($12_8$--$11_8$), or with a lower optical depth, e.g., CH$_3^{13}$CN 
($12_2$--$11_2$), are subject to limited signal-to-noise ratios and barely show the 
NW-SE velocity gradient. From the first moment maps of CH$_3$OH 
(15$_{4,11}$--16$_{3,13}$)$E$ and CH$_3$CN ($12_7$--$11_7$), the velocity gradient is
approximately along a position angle (PA) of $140^{\circ}$, which is adopted for the 
cut of the $P$-$V$ diagrams. The vibrational CH$_3$CN ($12_1$--$11_1$) line has the 
highest excitation and lowest signal-to-noise ratio among the three lines shown here.
The direction of the velocity gradient in its first-moment map is less clear, but an 
overall NW-SE gradient is still discernable and the $P$-$V$ pattern is consistent 
with those of the other two lines. Compared to the orientation of the biconical NE-SW
outflow, the CH$_3$OH and CH$_3$CN velocity gradient is approximately perpendicular 
to the outflow axis, and is apparently arising from a disk-like rotation. However, 
neither the dust continuum nor the molecular line maps of the MM1 condensation shows 
a flattened morphology. This is very likely due to insufficient angular resolutions 
($\gtrsim0.\!''5$) of the observations.

\subsection{Infall Signatures Seen in CN ($N$=2--1)} \label{res_CN}
\begin{figure}
\epsscale{.9} \plotone{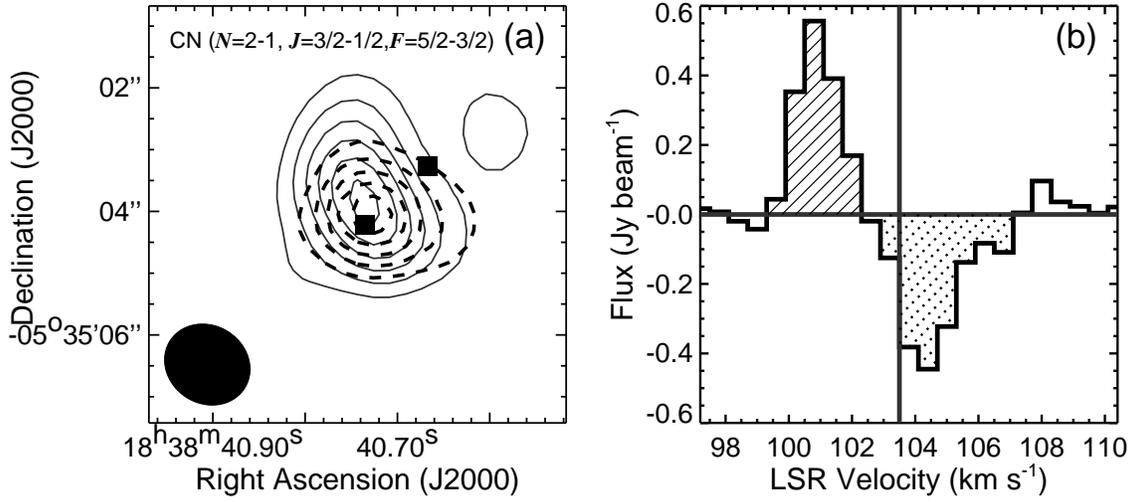} \caption{(a) Solid and dashed contours show the emission integrated from 96.6 to 102.0 km\,s$^{-1}$ and the absorption integrated from 103.2 to 106.8 km\,s$^{-1}$, respectively, in CN ($N$=2--1, $J$=3/2--1/2, $F$=5/2--3/2). The starting and spacing contour levels are 54 mJy\,km\,s$^{-1}$ for the emission and $-63$ mJy\,km\,s$^{-1}$ for the absorption. (b) The spectrum of the same CN transition, extracted from MM1. A vertical line marks the systemic velocity. The velocity channels integrated to show the emission and absorption in (a) are shaded with slashes and dots, respectively.
\label{absorp}}
\end{figure}

Several CN ($N$=2--1) hyperfine lines are detected toward MM1. The brightest three,
i.e., the $J$=5/2--3/2, $F$=5/2--3/2 (${\nu}_0$=226.87417 GHz), $J$=5/2--3/2,
$F$=7/2--5/2 (226.87474 GHz), and $J$=5/2--3/2, $F$=3/2--1/2 (226.87590 GHz) lines,
are heavily blended and form a triplet. Besides the triplet, the lines of
$J$=3/2--1/2, $F$=5/2--3/2 (226.65956 GHz) and $J$=3/2--1/2, $F$=3/2--1/2 (226.67931
GHz) are detected as well; these two lines do not suffer from line blending and the
brighter one, CN ($N$=2--1, $J$=3/2--1/2, $F$=5/2--3/2), is shown in Figure
\ref{absorp}. The CN ($N$=2--1) transitions have $E_{\rm up}\sim16$ K and 
$n_{\rm cr}$ of order $10^7$ cm$^{-3}$, probing relatively cold gas in a dense 
envelope around MM1. All the detected CN lines show emission at blueshifted 
velocities and absorption at redshifted velocities, i.e., an inverse P-Cygni profile.
Such a profile is often interpreted to indicate an infall of matter, whereas the 
absorption mostly comes from the gas on the near side of a central source and the 
emission is attributed to the gas in the far side 
\citep[e.g.,][]{Zapata08,Wu09,Qiu11b}. Therefore, the CN lines probe an infalling 
envelope around MM1. In Figure \ref{absorp}, the peaking velocities of the emission 
and absorption are slightly asymmetric with respective to the systemic velocity. It 
could be in part due to the uncertainty in determining the systemic velocity, which 
is estimated to be 103.5 km\,s$^{-1}$ in our data but was suggested to be 102.3 
km\,s$^{-1}$ based on single-dish NH$_3$ observations \citep{Molinari96}.

\section{Discussion} \label{dis}
\subsection{An accreting star of $\sim$10 $M_{\odot}$}
MM1 has a gas temperature of 320 K and molecular chemistry characteristic of a hot
molecular core (HMC). At least one high-mass protostar has formed in MM1. On the
other hand, only very faint and spatially unresolved radio continuum emission is
detected. The emission can be arising from an ionized jet or wind, or from a
hypercompact H {\scriptsize II} region. In the latter case, the central ionizing star
has an equivalent spectral type of B0.5, corresponding to a stellar mass, $M_{\ast}$,
of $12~M_{\odot}$ \citep{Schaller92}.
The mass of the outflowing gas can also shed light on the stellar mass of the
protostar. Assuming that the molecular outflow is momentum driven by an underlying
wind, the accumulated mass ejected to the wind is $M_{\rm out}v_{\rm out}/v_{\rm w}$,
where $M_{\rm out}$ is the outflow mass, $v_{\rm out}$ the outflow velocity, and
$v_{\rm w}$ the wind speed. The mass loss rate to the wind is a fraction, $f$, of the
mass infall rate, thus the mass of the forming star can be expressed as $M_{\rm
out}v_{\rm out}/v_{\rm w}\cdot(1-f)/f$ \citep[see also][]{Lada96}. The outflow mass
is calculated from the CO line wings ($\leq$100.5 km\,s$^{-1}$ and $\geq$106.5
km\,s$^{-1}$), assuming an excitation temperature of 30 K and a CO abundance of
$10^{-4}$. To correct for the optical depth effect, we compare the CO and $^{13}$CO
(2--1) fluxes and adopt a C-to-$^{13}$C abundance ratio of 37 \citep{Wilson94}. A
more detailed description of the calculation procedure can be found in Qiu et al.
(2009). We derive an outflow mass of $54~M_{\odot}$ with the combined SMA and IRAM
30m data. The maximum velocity of the outflow observed in CO (2--1) is about 36
km\,s$^{-1}$. Adopting a wind speed of 500~km\,s$^{-1}$ \citep[e.g.,][]{Marti98} and
$f=1/3$ \citep{Tomisaka98,Shu00}, the accumulated stellar mass is about
$8~M_{\odot}$, in agreement with the above estimate.

The stellar luminosity of a $\sim$10 $M_{\odot}$ star at the Zero-Age-Main-Sequence
(ZAMS) is $\sim$10$^4~L_{\odot}$, only a small fraction of the total luminosity 
observed in IRAS 18360-0537 ($\sim$10$^5~L_{\odot}$). The majority of the observed 
luminosity presumably comes from accretion shocks. The accretion luminosity can be 
estimated from $GM_{\ast}\dot{M}/R_{\ast}$, where $\dot{M}$ is the mass flux of the 
accretion onto the star and $R_{\ast}$ the radius of the star. Again, under the 
assumption of momentum conservation between the molecular outflow and the underlying 
wind, $\dot{M}$ can be inferred from the mass outflow rate. Without correcting for an
unknown inclination angle, the dynamical timescale of the outflow, i.e., a half of 
the outflow extension ($\sim$0.3 pc) divided by the terminal velocity, is about 
$8\times10^3$ yr. The mass outflow rate is then 
$7\times10^{-3}~M_{\odot}$\,yr$^{-1}$, leading to a mass infall rate of 
$1.5\times10^{-3}~M_{\odot}$\,yr$^{-1}$ and 
$\dot{M}\sim1.0\times10^{-3}~M_{\odot}$\,yr$^{-1}$. If we adopt a radius of 
$4R_{\odot}$ for a 10 $M_{\odot}$ star at the ZAMS \citep{Panagia73,Schaller92}, the 
accretion luminosity amounts to $8\times10^4~L_{\odot}$, which does appear to 
dominate the total luminosity.

Alternatively, according to the model of Hosokawa et al. (2010), massive protostars
with disk accretion rates of $10^{-3}~M_{\odot}$\,yr$^{-1}$ attain a total
luminosity of $\sim$10$^5~L_{\odot}$ as $M_{\ast}$ increases to $\sim$15 $M_{\odot}$, 
and reach the ZAMS for $M_{\ast}\simeq30~M_{\odot}$. In this model, the stellar 
radius swells to tens of $R_{\odot}$ as $M_{\ast}$ reaches $\sim10~M_{\odot}$, so the
accretion luminosity is only of order $10^4~L_{\odot}$, while the stellar radiation, 
which comes from the release of the gravitational energy, is dominating the total 
luminosity. The large radius of the protostar also reduces the effective temperature 
and the UV luminosity, thus the faint radio continuum emission is expected to be 
arising from an ionized jet or wind. 

Nevertheless, the protostar embedded in MM1 appears to have a current mass of $\sim$10 
$M_{\odot}$ and is in an active accretion phase.

\subsection{Rotation and Infall} \label{dis2}
\begin{figure}
\epsscale{.9} \plotone{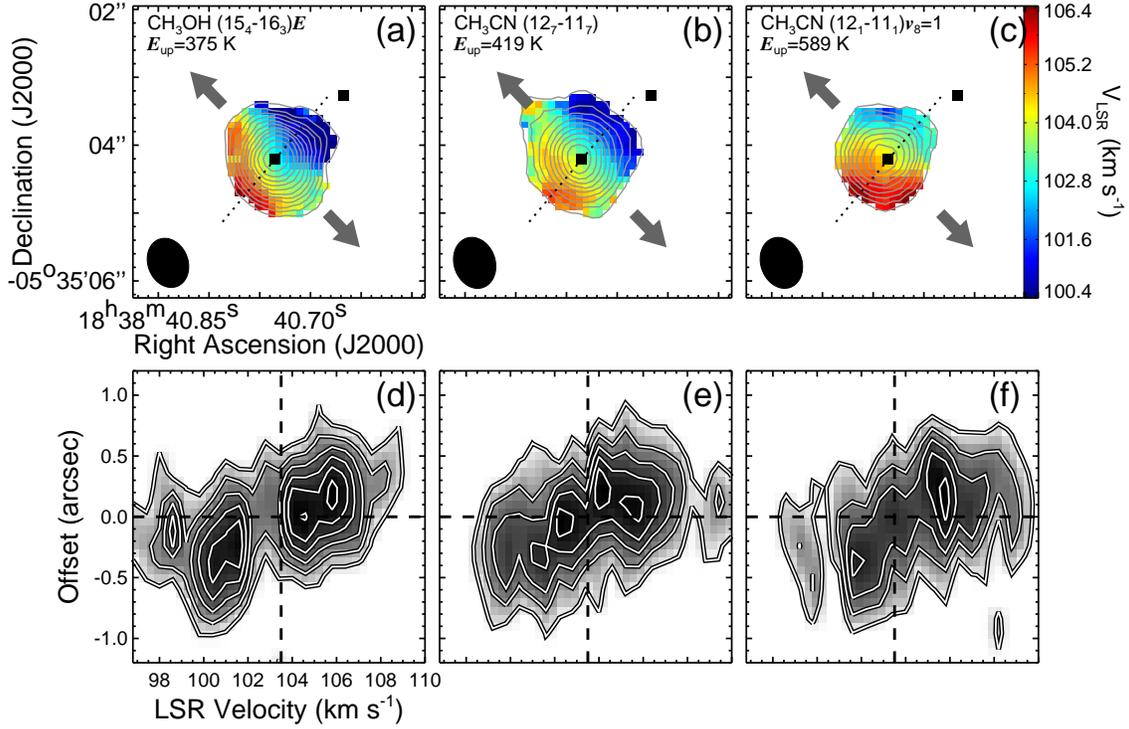} \caption{(a): The color image shows the first moment map of the CH$_3$OH (15$_{4,11}$--16$_{3,13}$)$E$ emission. The gray contours show the zeroth moment map starting from 1.0 Jy\,km\,s$^{-1}$ and increasing in steps of 0.8 Jy\,km\,s$^{-1}$. A dotted line marks the $P$-$V$ cut at PA=$140^{\circ}$. Two arrows are the same as those in Figure \ref{outflow}, indicating the outflow orientation. (b)--(c) Same as (a), but for CH$_3$CN ($12_7$--$11_7$) and vibrational CH$_3$CN ($12_1$--$11_1$). The contour levels in panel (b) start from 0.6 Jy\,km\,s$^{-1}$ and increase in steps of 0.8 Jy\,km\,s$^{-1}$, and in panel (c) start from 1.3 Jy\,km\,s$^{-1}$ and increase in steps of 0.7 Jy\,km\,s$^{-1}$. The color wedge on the right of panel (c) indicates the velocity scale of the first moment maps. (d)--(f) $P$-$V$ diagrams shown in both gray scale and contours. The contour levels all start from 0.21 Jy\,beam$^{-1}$, and increase in steps of 0.35, 0.21, and 0.28 Jy\,beam$^{-1}$ for CH$_3$OH (15$_{4,11}$--16$_{3,13}$)$E$ (d), CH$_3$CN ($12_7$--$11_7$) (e), and vibrational CH$_3$CN ($12_1$--$11_1$) (f), respectively. The horizontal line marks the peak position of MM1, which is taken to be the zero offset. The vertical line denotes the systemic velocity.
\label{rotation}}
\end{figure}

Is the accretion onto the central high-mass protostar mediated by a rotating disk?
The observed CO and SiO outflow supports a positive answer since a biconical outflow
is expected to originate from an accretion disk in low-mass star formation
\citep[see][for a detailed comparison between wide-angle outflows in low-mass and
high-mass star formation]{Qiu09b}. The high-excitation CH$_3$OH and CH$_3$CN lines
observed at sub-arcsecond resolution provide direct evidence for the rotation of the
HMC. There seem to be hints of a Keplerian-like motion in the $P$-$V$ diagrams of the
high-excitation lines. For example, in the upper right quadrant of Figure
\ref{rotation}(e), the maximum velocity increases with the decreasing offset; in the
lower left quadrant of Figure \ref{rotation}(d), a secondary peak around 98.5
km\,s$^{-1}$ could be attributed to a Keplerian-like motion as well.

\begin{figure}
\epsscale{.9} \plotone{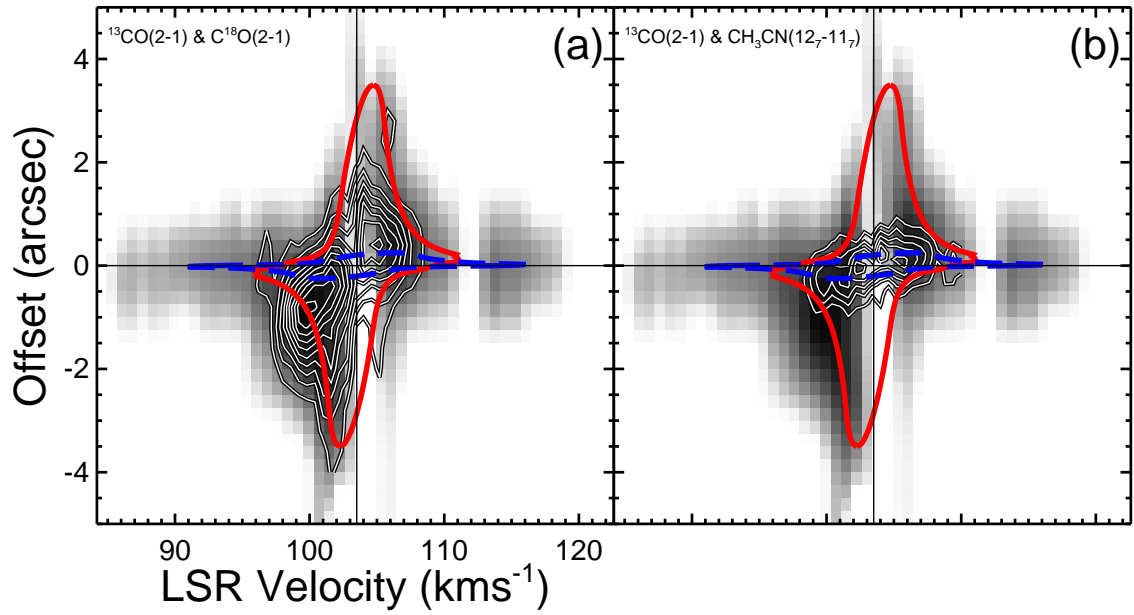} \caption{The $^{13}$CO (2--1) $P$-$V$ diagram in gray scale, overlaid with those in C$^{18}$O (2--1) (a) and CH$_3$CN ($12_7$-$11_7$) (b) in contours. The C$^{18}$O starting and spacing contour levels are 0.21 Jy\,beam$^{-1}$. In each panel, the red solid curve shows the model of a free-falling and Keprelian-like rotating envelope, and the blue dashed curve represents the same model but with scaled down parameters (see Section \ref{dis2} for more details of the models).
\label{rotcur}}
\end{figure}

A more detailed investigation of the rotation curve of the HMC awaits future 
observations with greatly improved resolution and sensitivity. Here we look into the 
lines from more abundant molecular species, i.e., C$^{18}$O and $^{13}$CO, which may 
help to deliver a more complete view of the velocity field for the dense gas on a 
larger scale \citep[e.g.,][]{Cesaroni11}. $^{13}$CO (2--1) can be contaminated by the 
outflow, but from Figure \ref{co_13co}, the emission in the close vicinity of MM1 is 
dominated by the dense envelope. In Figure \ref{rotcur}(a) we plot the $^{13}$CO and 
C$^{18}$O $P$-$V$ diagrams, cut along the PA of $140^{\circ}$. We note that the 
$^{13}$CO and C$^{18}$O emission arises from a region encompassing both MM1 and MM2. 
However, from the $P$-$V$ diagrams the velocity field cannot be ascribed to an 
orbiting motion of a binary. In particular, the high-excitation lines shown in Figure
\ref{rotation} are not detected toward MM2; these lines show a clear velocity 
gradient across MM1. In Figure \ref{rotcur}, the CH$_3$CN, C$^{18}$O, and $^{13}$CO 
emission seems to be arising from different layers of a coherent structure centered 
at MM1, whereas MM2 is likely a fragment of this structure. Despite the missing flux 
closely around the cloud velocity, the $^{13}$CO $P$-$V$ diagram shows a concave 
outer edge and is slightly skewed in a sense consistent with the velocity gradient 
seen in CH$_3$OH and CH$_3$CN. A similar pattern is discernible in the C$^{18}$O 
$P$-$V$ diagram, though the emission has a lower velocity dispersion and a smaller 
spatial extent compared to the $^{13}$CO emission. Such a $P$-$V$ pattern may 
characterize a combination of infall and rotation \citep{Cesaroni11,Tobin12}. To 
corroborate this interpretation, we overlay in Figure \ref{rotcur} a free-fall and 
Keplerian-like rotation model which is described by Cesaroni et al. (2011). Since the
C$^{18}$O and $^{13}$CO emission mostly traces outer layers around the HMC, we set a 
central dynamical mass of 25 $M_{\odot}$, which is approximately the combined mass of
the HMC and the embedded high-mass protostar, and an inner truncation radius of 
$0.\!''25$ (0.008 pc), which is approximately the deconvolved radius of the CH$_3$OH 
and CH$_3$CN emission; the outer radius is set to be $3.\!''5$ (0.1 pc), roughly the 
radius of the C$^{18}$O and $^{13}$CO emission. The constructed model (solid curve) 
reasonably fits the C$^{18}$O and $^{13}$CO $P$-$V$ diagrams except the high velocity
tail close to the zero offset seen in $^{13}$CO. That high velocity tail can be 
partly ascribed to the inner rotating and infalling gas, as illustrated by a 
scaled-down model (dashed curve in Figure \ref{rotcur}) with a central mass of 10 
$M_{\odot}$ and outer and inner radii of $0.\!''25$ and $0.\!''032$ (200 AU), 
respectively, but it can also be arising from the outflowing gas. Overall the bulk 
$^{13}$CO and C$^{18}$O emission delineates an infalling and rotating envelope around
the HMC.

To assess the envelope mass, we performed a two-component Gaussian fitting to the
central compact part of the dust ridge with the MIRIAD task IMFIT, and derived a flux
0.96 Jy for the component encompassing MM1. Subtracting the HMC contribution (0.76
Jy, see Section \ref{res_cont}), the dust continuum flux from the cool envelope
amounts to 0.2 Jy, resulting in a gas mass of $42~M_{\odot}$ for $\beta=0.82$ and an
adopted temperature of 30 K \citep{Molinari96}. From the fitting, the MM2 component
has a flux of 0.2 Jy, consistent with the measurement based on Figure \ref{cont}(b)
(Section \ref{res_cont}). Hence, the total mass of the central compact part of the
dust ridge reaches $80~M_{\odot}$, whereas the mass of the envelope traced by the
$^{13}$CO and C$^{18}$O emission amounts to $67~M_{\odot}$. Such a massive envelope
is presumably gravitationally unstable. Indeed, MM2 is likely the consequence of the
fragmentation of the envelope. However, the HMC itself has a gas mass
($13~M_{\odot}$) comparable to the mass of the accreting protostar ($10~M_{\odot}$),
thus the latter can have an appreciable stabilizing effect on the former. Cesaroni et
al. (2007) suggest that a rotational structure of a mass comparable to that of the
central protostar, which is in turn less than $20~M_{\odot}$, resembles an accretion
disk rather than a transient toroid. In this sense, we may expect that as the central
protostar continues to accrete and increase its mass, the rotating HMC serves as an
accretion disk, and finally dissipates by enhanced UV radiation
\citep[e.g., by photoevaporation,][]{Hollenbach94} rather than by gravitational
fragmentation.

The infall motion of the envelope around the HMC is not only appreciable in the
C$^{18}$O and $^{13}$CO $P$-$V$ diagrams, but also more unambiguously evidenced by
the CN hyperfine lines with a profound inverse P-Cygni profile. The mass infall rate
can be estimated from $4{\pi}r^2{\rho}V_{\rm in}$, where the infall velocity $V_{\rm
in}$ is about 1.5 km\,s$^{-1}$ according to the CN ($N$=2--1) profiles (e.g., Figure
\ref{absorp}), $r$ the radius of the envelope, and $\rho$ the mass density.
Considering that the CN lines were imaged with a synthesized beam
($1.\!''4\times1.\!''3$) similar to that of the dust continuum map in Figure
\ref{cont}(a) ($1.\!''6\times1.\!''2$), we estimate the radius and density based on
the above two-component Gaussian fitting. For the component encompassing MM1,
$r\simeq0.\!''9$ (5700 AU) from the fitting, and the averaged number density is
$8\times10^6$~cm$^{-3}$ for a gas mass of $42~M_{\odot}$. The mass infall rate is
then found to be about $1.5\times10^{-3}~M_{\odot}$\,yr$^{-1}$. From the model shown
in Figure \ref{rotcur}, the free-fall velocity reaches 3 km\,s$^{-1}$ at
$r\sim0.\!''9$, a factor of two higher that inferred from the CN profiles. Such a
discrepancy can be ascribed to the deviation of the infall motion from the modeled
free-fall, the uncertainty of the systemic velocity (see Section \ref{res_CN}), or an
underestimate of the radius from the two-component fitting. Nevertheless, since the
(free-fall plus Keplerian-like rotation) model parameters are not stringently
constrained and the CN profiles provide an averaged estimate, the discrepancy is not
significant. On the other hand, as discussed above, the mass outflow rate implies a
mass infall rate of $1.5\times10^{-3}~M_{\odot}$\,yr$^{-1}$. Though both are subject
to large uncertainties, the two independent estimates of the mass infall rate appear
to agree with each other, suggesting that the envelope infall is feeding the
accretion onto one protostar, i.e., the $\sim$10 $M_{\odot}$ star embedded in MM1.
Provided its high accretion rate and the massive gas reservoir available from the
envelope, it is reasonable to expect that the star would significantly increase its
mass and develop into an O star.

\section{Summary} \label{summ}
We have presented high-angular-resolution dust continuum and molecular line
observations toward IRAS 18360-0537, a $\sim$10$^5~L_{\odot}$ star-forming region.
The region shows a 0.5~pc dust ridge, harboring two massive condensations, MM1 and
MM2, in the central part. MM1, a HMC with a gas temperature of 320 K and a mass of
$13~M_{\odot}$, is embedded within an infalling envelope, and appears to be rotating
about the axis of a bipolar outflow.

The outflow is seen in CO (2--1) and SiO (5--4) and shows a biconical shape centered
at MM1. The gas mass of the outflow is estimated to be $54~M_{\odot}$, and the
dynamical timescale is $8\times10^3$ yr. The rotation of the HMC is inferred from
sub-arcsecond observations of high-excitation CH$_3$OH and CH$_3$CN lines, which show
a velocity gradient approximately perpendicular to the outflow axis. On the other
hand, the infall motion of the envelope around the HMC is probed by CN ($N$=2--1)
hyperfine lines, which exhibit a profound inverse P-Cygni profile. The estimated mass
infall rate is $1.5\times10^{-3}~M_{\odot}$\,yr$^{-1}$, comparable to that inferred
from the mass outflow rate. Further information on the kinematics of the gas around
the HMC is obtained from the C$^{18}$O and $^{13}$CO (2--1) lines, whose $P$-$V$
diagrams can be described with a model of a free-falling and Keplerian-like rotating
envelope. With these observations, we speculate that a protostar in MM1, which we
suggest to have a current mass of $\sim$10 $M_{\odot}$, will grow into an O star via
disk accretion and envelope infall, i.e., through a process similar to the standard
paradigm for low-mass star formation.


\clearpage

\clearpage

\clearpage

\begin{deluxetable}{ccccccccc}
\tabletypesize{\scriptsize}
\rotate
\tablecaption{Summary of SMA Observations \label{table1}}
\tablewidth{0pc}
\tablehead{ \colhead{} & \colhead{} & \multicolumn{2}{c}{Freq. Cove. (GHz)} & \colhead{} & \colhead{} & \colhead{} \\
\cline{3-4} \\[-2ex]
\colhead{Obs. Date} & \colhead{Array Conf.} & \colhead{LSB} & \colhead{USB} & \colhead{Spec. Res. (kHz)} &
\colhead{Bandpass Cal.} & \colhead{Gain Cal.} & \colhead{Flux Cal.} & \colhead{${\tau}_{\rm 225 GHz}$} \\
\colhead{(1)} & \colhead{(2)} & \colhead{(3)} & \colhead{(4)} & \colhead{(5)} & \colhead{(6)} & \colhead{(7)}
& \colhead{(8)} & \colhead{(9)} }

\startdata
2007 May 17 & VEX (6) & 219.3 -- 221.3 & 229.3 -- 231.3 & 406.25 & 3C273 &  J1743-038, J1911-201 & Vesta & 0.1 \\
2007 Jul. 08 & Comp. (8) & 219.3 -- 221.3 & 229.3 -- 231.3 & 406.25 & 3C273 & J1743-038, J1911-201 & Vesta & 0.07 \\
2008 May 21 & Comp. (8) & 215.4 -- 217.4 & 225.4 -- 227.4 & 812.5 & 3C273 & J1743-038, J1911-201 & Callisto & 0.15 \\
2008 Jul. 19 & EXT (6) & 215.4 -- 217.4 & 225.4 -- 227.4 & 406.25 & 3C273, 3C454.3 & J1743-038, J1911-201 & Uranus & 0.07 \\
\enddata
\tablecomments{Col.(1): Observing dates. Col.(2): Array configurations, with the Very Extended, Compact, and Extended configurations abbreviated to VEX, Comp., and EXT, respectively; the figure in parentheses denotes the number of antennas used for each observation. Col.(3)--(4): Approximate ranges of rest frequencies covered in the lower sideband (LSB) and upper sideband (USB). Col.(5): Spectral resolutions. Col.(6)--(8): Bandpass, time dependent gain, and absolute flux calibrators. Col.(9): The averaged 225 GHz opacity.}
\end{deluxetable}

\end{document}